\documentclass[prl,twocolumn,aps]{revtex4-1}
\usepackage{graphicx}
\usepackage{amsfonts}
\usepackage{amsmath,amssymb,amsthm,bm,multirow,longtable}
\usepackage{epsfig}
\usepackage{epstopdf}
\usepackage{comment}
\usepackage{color}
\newcommand{\nn}{\nonumber}
\renewcommand{\epsilon}{\varepsilon}

\newcommand{\be}{\begin{equation}}
\newcommand{\ee}{\end{equation}}
\newcommand{\bea}{\begin{eqnarray}}
\newcommand{\eea}{\end{eqnarray}}
\newcommand{\beq}{\begin{equation}}
\newcommand{\eeq}{\end{equation}}
\newcommand{\beqn}{\begin{eqnarray}}
\newcommand{\eeqn}{\end{eqnarray}}

\begin{document}
\title{Electromagnetic signatures of the chiral anomaly in Weyl semimetals}
\author{Edwin Barnes}%\email{efbarnes@vt.edu}
\author{J. J. Heremans}%\email{heremans@vt.edu}
\author{Djordje Minic}%\email{dminic@vt.edu}
\affiliation{Department of Physics, Virginia Tech, Blacksburg, VA 24061, U.S.A.}

\date{\today}
\begin{abstract}
Weyl semimetals are predicted to realize the three-dimensional axial anomaly first discussed in particle physics. The anomaly leads to unusual transport phenomena such as the chiral magnetic effect in which an applied magnetic field induces a current parallel to the field. Here we investigate diagnostics of the axial anomaly based on the fundamental equations of axion electrodynamics. We find that materials with Weyl nodes of opposite chirality and finite energy separation immersed in a uniform magnetic field exhibit an anomaly-induced oscillatory magnetic field with a period set by the chemical potential difference of the nodes. In the case where a chemical potential imbalance is created by applying parallel electric and magnetic fields, we find a suppression of the magnetic field component parallel to the electric field inside the material for rectangular samples, suggesting that the chiral magnetic current opposes this imbalance. For cylindrical geometries, we instead find an enhancement of this magnetic field component along with an anomaly-induced azimuthal component. We propose experiments to detect such magnetic signatures of the axial anomaly.

\end{abstract}

\maketitle

Dirac/Weyl semimetals are topological states of matter in which the 3D bulk contains Dirac or Weyl points protected by crystalline symmetries and near which the low-energy quasiparticles are linearly dispersing massless Dirac/Weyl fermions \cite{HerringPR1937,AbrikosovJETP1971,WanPRB2011,BurkovPRL2011,YoungPRL2012,ZyuzinPRB2012a}. While Dirac semimetals were theorized early on \cite{HerringPR1937,AbrikosovJETP1971}, Weyl semimetals were predicted only recently, with pyrochlore iridates such as Y${}_2$Ir${}_2$O${}_7$ as examples \cite{WanPRB2011}. The prediction that Weyl semimetals host exotic topological surface states distinct from those in other materials spurred extensive experimental efforts to confirm the existence of both Dirac and Weyl semimetal phases, leading to Dirac semimetal discoveries such as Na${}_3$Bi and Cd${}_3$As${}_2$ \cite{Wang_PRB13,LiuScience2014,BorisenkoPRL2014,LiuNatMater2014,NeupaneNatCommun2014}, and recent Weyl semimetal discoveries such as TaAs and NbAs \cite{HuangNatComm15,WengPRX2015,LvPRX2015,LvNP2015,XuScience2015b,XuNP15}. 

Theoretical works have shown that Dirac semimetals lie at the intersection of several types of topological states reachable by breaking symmetries including inversion, time-reversal, or crystal symmetries \cite{WanPRB2011,BurkovPRL2011,YoungPRL2012,ZyuzinPRB2012a,Wang_PRB13,WengPRX2015,HuangNatComm15}. For example, breaking crystal symmetries can cause Weyl nodes in the bulk to couple and open a band gap, which can produce a topological insulator \cite{Wang_PRB13}. The breaking of either time-reversal or inversion symmetry can result in a stable Weyl semimetal state which hosts topological Fermi arc states on the material's surface \cite{WanPRB2011,BurkovPRL2011,HosurPRL2012,Wang_PRB13,XuScience2015b,WengPRX2015,HuangNatComm15}. 

Weyl semimetals are also expected to show unusual phenomena associated with the three-dimensional axial anomaly \cite{ZyuzinPRB2012b,Chen_PRB13,Khaidukov_arxiv13,Goswami_PRB13}. Notable among these is the chiral magnetic effect (CME) in which the application of an external magnetic field $\vec{B}$ produces a current $\vec{j}\Vert\vec{B}$ \cite{SonPRL2012,Son_PRB13,Burkov_PRL14}. This effect has also been studied in the contexts of  quark-gluon plasmas \cite{Fukushima_PRD08, Nielsen_PLB83} and topological insulators \cite{Fu_PRB07, Teo_PRB08, Qi_PRB08, Hasan_RMP10, Li_NP10,Kim_PRL13}. Transport measurements in Weyl semimetals revealed a negative longitudinal magnetoresistance as a diagnostic for the CME \cite{He_PRL14,LiangNatMater2015,Zhang_NC16,LiNatPhys2016}. Debate exists whether this transport signature necessarily implies the CME \cite{Goswami_PRB15}, and in this Letter we thus develop a diagnostic based on magnetic properties, which finds a parallel in the development of the London penetration depth in superconductors.

The approach is based on self-consistent solutions to Maxwell's equations in the presence of anomalous chiral currents. Even for currents derived in the limit of linear response, the electromagnetic fields should be obtained self-consistently as becomes clear when starting with the fundamental equations of axion electrodynamics---the relevant effective field theoretic description when the axial anomaly is present. We obtain such solutions for systems in which the chiral chemical potential is either intrinsic or induced by parallel electric and magnetic fields. In both cases, we find that the solutions exhibit detectable signatures due to the presence of the axial anomaly.

The defining characteristic of axion electrodynamics is captured by the
axionic term in the action \cite{Wilczek_PRL87} (in addition to the canonical  $\frac{1 }{4} \int d^3 x dt ({\vec{E}}^2 - {\vec{B}}^2)$ term), 
\be
S_{\theta} = \frac{\alpha }{4 \pi^2} \int d^3 x dt\theta \vec{E} \cdot \vec{B},
\ee
where $\vec{E}$ is the electric field, and $\alpha = \frac{e^2}{\hbar c}$ is the fine structure constant. 
The parameter $\theta$, when dependent on space and time, is called the axion
field in the particle physics context. The relevant equations of axion electrodynamics are well known \cite{Goswami_PRB13,Sikivie_PLB84,Huang_PRD85,Wilczek_PRL87,Carroll_PRD90} 
\be
\nabla \cdot   \vec{E} = \rho - \kappa \nabla \theta \cdot \vec{B},
\quad 
\nabla \times \vec{E} = - \frac{\partial \vec{B}}{\partial t},\label{axion1}
\ee
together with
\be
\nabla \cdot   \vec{B} = 0,
\quad
\nabla \times \vec{B} = \frac{\partial \vec{E}}{\partial t} + \vec{j} + \kappa \left(\frac{\partial \theta}{\partial t} \vec{B} + \nabla \theta \times \vec{E}\right).\label{axion2}
\ee
Note that when the axion profile is linear in time, $\theta \sim t$, one obtains for stationary fields that $\nabla \times \vec{B}=\zeta\vec{B}$ with constant $\zeta$, a result known as the Beltrami equation and which is well known \footnote{We thank Roman Buniy and Tom Kephart for discussions about this connection.} in the context of 
plasma physics \cite{Beltrami_RRIL1889,Buniy_AP13,Gorbar_PRB13,Khaidukov_arxiv13}.

We proceed with a brief review of the CME as a physical consequence of the $U(1)$ chiral anomaly \cite{Fukushima_PRD08}.
Consider the minimal coupling of the carriers of $\vec{j}$ such as chiral fermions to an external $U(1)$ gauge field.
First we rewrite the axion term in a covariant way
$
\theta(x,t) \vec{E} \cdot \vec{B} \equiv \theta(x,t) F^{\mu \nu} \tilde{F}_{\mu \nu}.
$
Then we perform a $U(1)$ rotation in the minimal coupling of fermions to
the gauge field in order to obtain
$
\partial_\mu \theta \bar{\psi} \gamma^{\mu} \gamma^{5} \psi.
$
In particular if we concentrate on the $0$th (temporal) component we get
$
\mu_5 \bar{\psi} \gamma^{0} \gamma^{5} \psi,
$
where the chiral chemical potential $\mu_5$ is given as
$
\mu_5 \equiv \partial_0 \theta.
$
The energy spectrum of the free Dirac equation in the presence of a
chiral chemical potential is for the massless modes (assuming 1D)
$
E_{R\pm} = \pm p_3 - \mu_5, \quad E_{L\pm} = \pm p_3 + \mu_5,
$
where $\pm$ represents the spin in the $z$ direction and
$R, L$ stand for the right and left chirality.
If the chiral chemical potential $\mu_5$ is positive, a net chirality is created, thus
lifting the degeneracy between the R and L modes.
$\vec{B}$ lifts the degeneracy in spin, depending on the charge of the particle.
Therefore particles with the right-handed helicity will move opposite to the antiparticles (holes)
with the right-handed helicity, thus creating $\vec{j}\Vert\vec{B}$.
This is the CME \cite{Fukushima_PRD08}.
The physics underlying this reasoning harks back to the original papers on chiral anomalies \cite{Adler_PR69, Bell_NuovoCim69}.
The total current is given as the volume integral, 
$
J^{\mu} = \int d^3 x j^{\mu} (x),
$
where the current density is given as the expectation value, 
$
j^{\mu} = e \langle \bar{\psi} (x) \gamma^{\mu} \psi(x) \rangle,
$
and where the fermion field can be written in terms of the left- and right-handed components
$\psi = (\varphi_L, \varphi_R)^T$, so that,  
$
j^{\mu} = e \langle \varphi_R^{\dagger}(x) \sigma^{\mu} \varphi_R(x) \rangle
+  e \langle \varphi_L^{\dagger}(x) \bar{\sigma}^{\mu} \varphi_L(x) \rangle.
$
Here
$\sigma^{\mu} \equiv (1, \sigma^i)$ and $\bar{\sigma}^{\mu} = (1, -\sigma^i)$,
where $\sigma^i$ are the canonical Pauli matrices.

As pointed out in \cite{Fukushima_PRD08}, one path to the CME in the context of quark-gluon plasmas uses the argument based on energy
balance as presented originally in \cite{Nielsen_PLB83}.
One considers a situation with $\vec{E}$ and $\vec{B}$ 
in the presence of a chiral chemical potential $\mu_5$ and relates the
work performed by $\vec{j}$ in $\vec{E}$ to the energy penalty related to the
chirality change (given essentially by the volume integral over $\vec{E} \cdot \vec{B}$), i.e.
$
\int d^3x \vec{j} \cdot \vec{E} = - \frac{e^2 \mu_5}{2 \pi^2} \int d^3x \vec{E} \cdot \vec{B}.
$
Therefore $\vec{j}$ is proportional to $\vec{B}$ (even in the $\vec{E} \to 0$ limit) 
$
\vec{j} = - \frac{e^2 \mu_5}{2 \pi^2} \vec{B}
$
which is the defining expression for the CME.

The final equation (in SI units, with constants recovered to facilitate evaluation), $\vec{j} = -(\tfrac{e^2}{\hbar^2}) (\tfrac{\mu_5}{2\pi^2}) \vec{B}$, does not depend on covariance and
thus it can be realized in many-body systems, such as Weyl semimetals. In this context, $\mu_5$ should be interpreted as the energy separation of bulk Weyl nodes, $\Delta\epsilon$ \cite{ZyuzinPRB2012b,Chen_PRB13}. In the case of Weyl semimetals, one must take care to show that nonlinearities in the dispersion do not remove the anomaly by solving a quantum kinetic equation. This approach has been used to show that the chiral magnetic current is not an equilibrium current, but $\vec{j}\sim\vec{B}$ still holds at arbitrarily low frequencies \cite{Vazifeh_PRL13,Chen_PRB13,Goswami_PRB13}. A second important difference in the Weyl semimetal context is that the axion field also depends on the Weyl node momentum separation: $\theta\sim \Delta \vec p\cdot\vec x$, so that the last term in Eq.~\eqref{axion2} gives rise to an anomalous Hall effect \cite{ZyuzinPRB2012b}. We thus have the well-known result \cite{ZyuzinPRB2012b,Vazifeh_PRL13,Chen_PRB13,Goswami_PRB13} (in SI units, with $\mu_0 = 4 \pi\times10^{-7}\, \mathrm{kg\, m/C^2}$) 
\be
\vec{\nabla} \times \vec{B} =\mu_0\vec{j}= -\frac{\mu_0e^2}{2\pi^2\hbar^2}\Delta\epsilon \vec{B}+\frac{\mu_0e^2}{2\pi^2\hbar^2}\Delta\vec p\times\vec E.
\label{b3def2}
\ee
While our main focus here is on Weyl semimetals, future work may include generalization of our results to other quasi-relativistic materials and to high energy systems such as quark-gluon plasmas.

As a first example of the consequences of Eq.~\eqref{b3def2}, we consider a semi-infinite slab of Weyl semimetal occupying $z\ge0$. Suppose that $\vec{B}_\perp(z=0)=0$ and $\vec{B}_\parallel(0)=B_0\hat y$ is constant, and that $\vec E=0$; the solution to Eq.~\eqref{b3def2} is $\vec{B}=B_0[\hat y\cos(z/\lambda)-\hat x\sin(z/\lambda)]$ with $\lambda=2\pi^2\hbar^2/\mu_0\Delta\epsilon e^2$. Incidentally, this solution is a special case of the more general solution in plasma physics known as the Arnold-Beltrami-Childress flow \cite{Arnold_Springer98}. In the context of Weyl semimetals, we see that a consequence of the chiral anomaly is that inside the slab, $\vec{B}$ forms a standing wave with wavelength $\lambda\sim 1/\Delta\epsilon$.

In the case of time-dependent fields where $\vec{E} \sim e^{i \omega t}$ and  $\vec{B} \sim e^{i \omega t}$ and for constant $\dot\theta$, solutions to Eqs.~\eqref{axion1} and \eqref{axion2} can be obtained by requiring that both $\vec{E}$ and $\vec{B}$ satisfy Beltrami equations:
$\vec{\nabla} \times \vec{E} = \zeta \vec{E}$ and $\vec{\nabla} \times \vec{B} = \zeta \vec{B}$ with $2 \zeta = 1/\lambda - \sqrt{1/\lambda^2 + 4 \omega^2}$. Note that here we have chosen the minus sign in front of the square root to ensure that in the $\omega\to0$ limit, the wave vector of the magnetic field goes to zero faster than the frequency, $q\sim\zeta\to-\lambda\omega^2\ll\omega$, as is necessary to ensure the existence of the CME as a non-equilibrium transport phenomenon \cite{Chen_PRB13,Chang_PRB15,Ma_PRB15,Zhong_PRL16}. Such Beltrami-type solutions describing the pure CME are only valid if $\Delta\vec p\cdot\vec B=0$ and if the last term in Eq.~\eqref{b3def2} can be neglected. Since for such solutions $\vec E=-i\zeta^{-1}\omega\vec B$, the latter condition requires $\zeta^2\gg \omega\Delta p/\lambda\Delta\varepsilon$, which can be satisfied if we take $\omega\gg\Delta p/\lambda\Delta\varepsilon$. We must also ensure that $\omega$ is below the threshold for optical absorption: $\omega<\Delta\varepsilon/\hbar$ \cite{Chang_PRB15,Ma_PRB15,Zhong_PRL16}, which in turn gives a constraint on the Weyl node momentum and energy separations: $\Delta p/\Delta\varepsilon\ll2\pi^2\hbar/\mu_0e^2\approx1350$. We can then obtain Beltrami-type solutions within this frequency window by starting from static solutions such as the above semi-infinite slab solution and taking $\lambda\to\zeta^{-1}$. (The time-dependent solutions of axion electrodynamics have been considered in \cite{Fujita_arxiv16}.)

\begin{figure}
\includegraphics[width=\columnwidth]{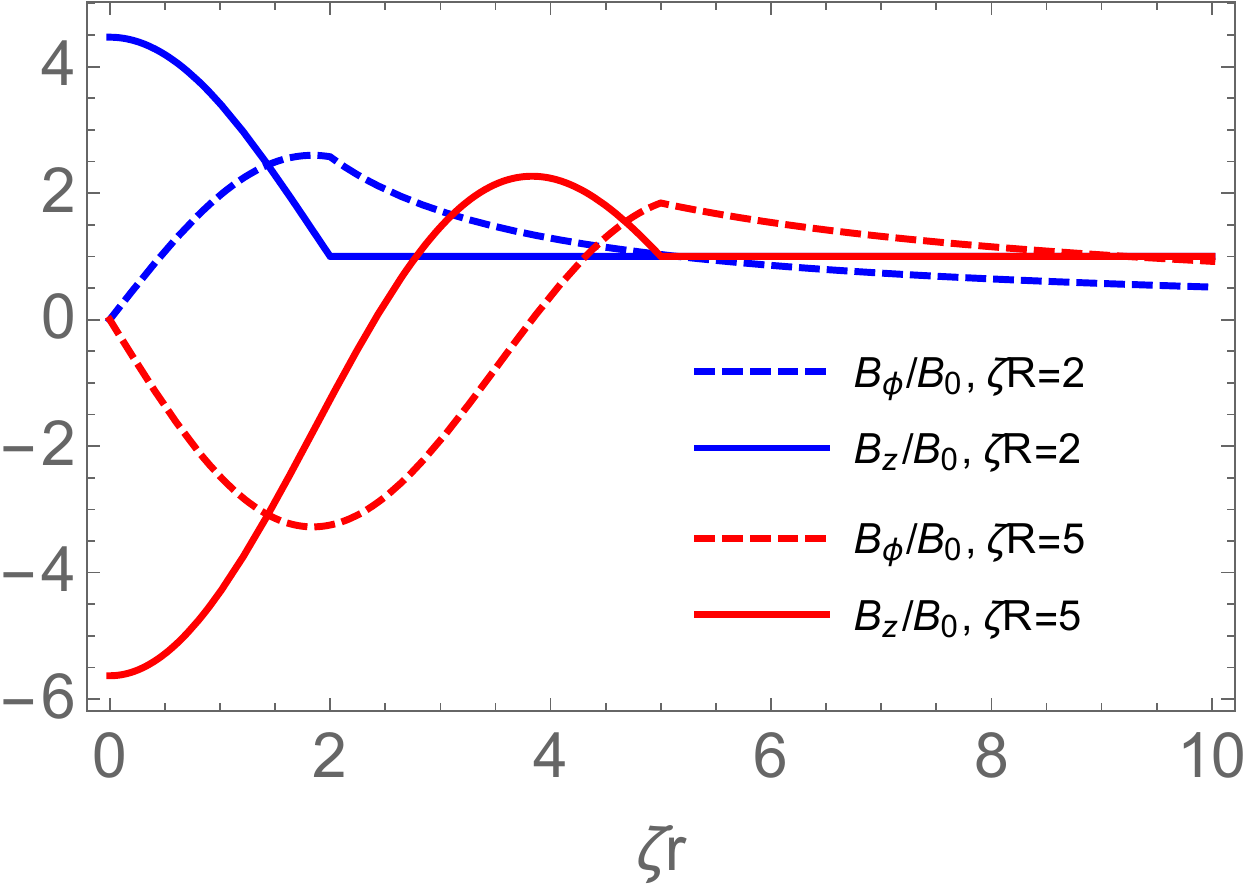}
\caption{\label{fig:cme_cyl} Components of $\vec{B}$ inside and outside a long cylinder of radius $R$ for $\zeta R=2,5$ subject to an applied field $B_0\hat z$.}
\end{figure}

Detectable signatures of axion electrodynamics can be obtained by solving the Beltrami equation in systems with finite dimensions. Consider a long cylinder of Weyl semimetal with radius $R$ and axis along $\hat z$ immersed in a constant magnetic field, $B_0\hat z$. We work in cylindrical coordinates ($r,\phi,z$) and require $\vec{B}$ to be independent of $\phi$ and $z$. The Beltrami equation reduces to
\be
\frac{1}{r}\frac{d}{dr}\left(r\frac{dB_z}{dr}\right)+\zeta^2B_z=0, \quad B_\phi=-\frac{1}{\zeta} \frac{d B_z}{dr},\label{helmholtz}
\ee
and $B_r=0$. These equations are easily solved to yield the field inside the cylinder: $B_z^{in}\sim J_0(\zeta r)$, $B_\phi^{in}\sim J_1(\zeta r)$. Outside the cylinder, $\vec{B}$ satisfies the ordinary Maxwell equations with a current source $\vec{j}\sim \vec{B}^{in}$. Using that $\vec{B}(r\gg R)=B_0\hat z$ and employing Stokes' theorem, we obtain the full solution inside and outside:
\bea
B_\phi&=&B_0\left[\frac{J_1(\zeta r)}{J_0(\zeta R)}\Theta(R-r)+\frac{RJ_1(\zeta R)}{rJ_0(\zeta R)}\Theta(r-R)\right],\nn\\
B_z&=&B_0\left[\frac{J_0(\zeta r)}{J_0(\zeta R)}\Theta(R-r)+\Theta(r-R)\right].\label{cylsol}
\eea
This solution is shown in Fig.~\ref{fig:cme_cyl}. The maximal value of $B_z$ occurs at the center of the cylinder, where $|B_z(0)|>|B_0|$. Outside the cylinder, the chiral anomaly gives rise to a $B_\phi$ which depends on $\Delta\epsilon$ and which varies quasi-periodically with the cylinder size $R$ at fixed $r$.

Although the CME has not yet been observed in Weyl semimetals where the energy separation between Weyl nodes persists in the absence of externally applied fields, experimental observations in a Dirac semimetal in which the degeneracy of Dirac nodes is lifted through the application of non-orthogonal $\vec{E}$ and $\vec{B}$ have been reported \cite{Zhang_NC16,Li_NC16,LiNatPhys2016,Xiong_Science15,Xiong_arxiv15b,Wang_PRB16} following theoretical predictions \cite{Gorbar_PRB13}. In this case, an anomalous current parallel to $\vec{B}$ is still generated \cite{Son_PRB13,Burkov_PRL14}, with a  magnitude depending on $\vec{E}\cdot\vec{B}$:
$
\vec{j}=\sigma_a (\vec{E}\cdot\vec{B})\vec{B}.
$
Unlike the pure CME relation, $\vec{j}\sim\vec{B}$, the above relation is not captured by axion electrodynamics and does not require time-dependent fields \cite{Zhong_PRL16}. However, combining this relation with Ampere's law, we still obtain a self-consistent equation for the local $\vec{B}$ in the material (SI units):
\be
\nabla\times\vec{B}=\mu_0\vec{j}=\mu_0\sigma_a(\vec{E}\cdot\vec{B})\vec{B}.\label{cmeDirac}
\ee
To illustrate consequences of Eq. \eqref{cmeDirac}, consider the case with constant $\vec{E}$ and $\vec{B}$ applied along the $y$-direction, $\vec{E}=E_0\hat y$, $\vec{B}_{ext}=B_0\hat y$, with the semimetal occupying the semi-infinite space $z\ge0$, and with boundary conditions $B_x(z=0)=0$, $B_y(z=0)=B_0$, $B_z(z=0)=0$. 
We look for solutions of the form $\vec{B}=B_x(z)\hat x+B_y(z)\hat y$, in which case Eq. \eqref{cmeDirac} reduces to:
\be
B_x'=\mu_0\sigma_a E_0B_y^2,\quad B_y'=-\mu_0\sigma_a E_0B_xB_y.\label{BxByeqns}
\ee
These two equations together imply 
$
B_x^2+B_y^2=B_0^2=const.
$
We can then parameterize the two components in terms of a new function, $\varphi(z)=\arctan(B_y/B_x)$, for which Eq. \eqref{BxByeqns} implies
$
\varphi'\csc\varphi=-\mu_0\sigma_a E_0B_0\equiv-\Lambda^{-1}.
$
This equation is readily solved:
$
\varphi=2\hbox{arctan}\left(e^{-z/\Lambda}\right),
$
with the constant of integration chosen to respect $B_y(0)=B_0$. We obtain for $\vec{B}$ inside the material: 
\be
\vec{B}=B_0\tanh(z/\Lambda)\hat x+B_0\hbox{sech}(z/\Lambda)\hat y.
\ee
Thus $B_y$ decays exponentially over a characteristic length $\Lambda$ into the bulk, while $B_x$ grows from $0$ to $B_0$ over a similar distance. In the bulk $\vec{B}$ thus becomes orthogonal to $\vec{E}$, in turn implying that $\vec{j}$ exists only near the surface. Maxwell's equations would thus seem to oppose the separation of Weyl nodes due to the chiral anomaly.

A similar effect persists even in the presence of an additional Ohmic current (conductivity $\sigma_0$):
$
\vec{j}=\mu_0\sigma_a (\vec{E}\cdot\vec{B})\vec{B}+\sigma_0\vec{E}.
$
With $\vec{E}=E_0\hat y$, Eq.~\eqref{BxByeqns} becomes
\be
B_x'=\mu_0\sigma_a E_0B_y^2+\mu_0\sigma_0E_0,\quad B_y'=-\mu_0\sigma_a E_0B_xB_y.\label{BxByeqns2}
\ee
Writing
$
B_x=|\vec{B}|\cos\varphi$, $ B_y=|\vec{B}|\sin\varphi,
$
Eqs.~\eqref{BxByeqns2} become
\be
\frac{d|\vec{B}|}{dz}=\mu_0\sigma_0E_0\cos\varphi,\quad
\varphi'\csc\varphi=-\frac{\mu_0\sigma_0 E_0}{|\vec{B}|}\left(\frac{\sigma_a}{\sigma_0}|\vec{B}|^2+1\right).\label{absBphi}
\ee
Unlike the previous case lacking the Ohmic term, here we note that not only the direction but also the magnitude of $\vec{B}$ varies with $z$. The second equation in \eqref{absBphi} can be integrated to yield 
$
\cos\varphi=\tanh\left[\mu_0 E_0\int_0^zdz'\left(\sigma_a|\vec{B}(z')|+\frac{\sigma_0}{|\vec{B}(z')|}\right)\right].
$
Combining this with the first equation in \eqref{absBphi} and writing $|\vec{B}(z)|=B_0\sqrt{h(\xi)}$, with $\xi\equiv\mu_0 E_0(\sigma_0/B_0+\sigma_a B_0)z$, yields
\be
h''(\xi)+\frac{\beta}{2}[h'(\xi)]^2-\frac{2\beta}{(1+\beta)^2} h(\xi)=\frac{2}{(1+\beta)^2},\label{heqn}
\ee
where $\beta\equiv\sigma_a B_0^2/\sigma_0$. We want to find solutions to \eqref{heqn} that obey the boundary conditions $h(0)=1$ (since $|\vec{B}(0)|=B_0$)  and $h'(0)=2\cos\varphi_0/(1+\beta)$, where $\varphi_0$ is the angle of the $\vec{B}$ relative to the $x$-axis at $z=0$. Given a solution to Eq.~\eqref{heqn} for the magnitude of $\vec{B}$, its orientation follows from
$
\cos\varphi=(1+\beta)\frac{d}{d\xi}\sqrt{h(\xi)}.
$
Before looking for solutions to Eq. \eqref{heqn} in the presence of both the Ohmic and anomalous currents, it is instructive to first consider the solution that arises in the absence of the anomalous current. Setting $\sigma_a=0$ in Eq.~\eqref{BxByeqns2}, we obtain
$
B_x=\mu_0\sigma_0E_0z+B_0\cos\varphi_0, \quad B_y=B_0\sin\varphi_0.
$
We see that the Ohmic current produces a nonzero $B_x$ that grows linearly with distance into the bulk, while $B_y$ remains constant.

\begin{figure}
\includegraphics[width=\columnwidth]{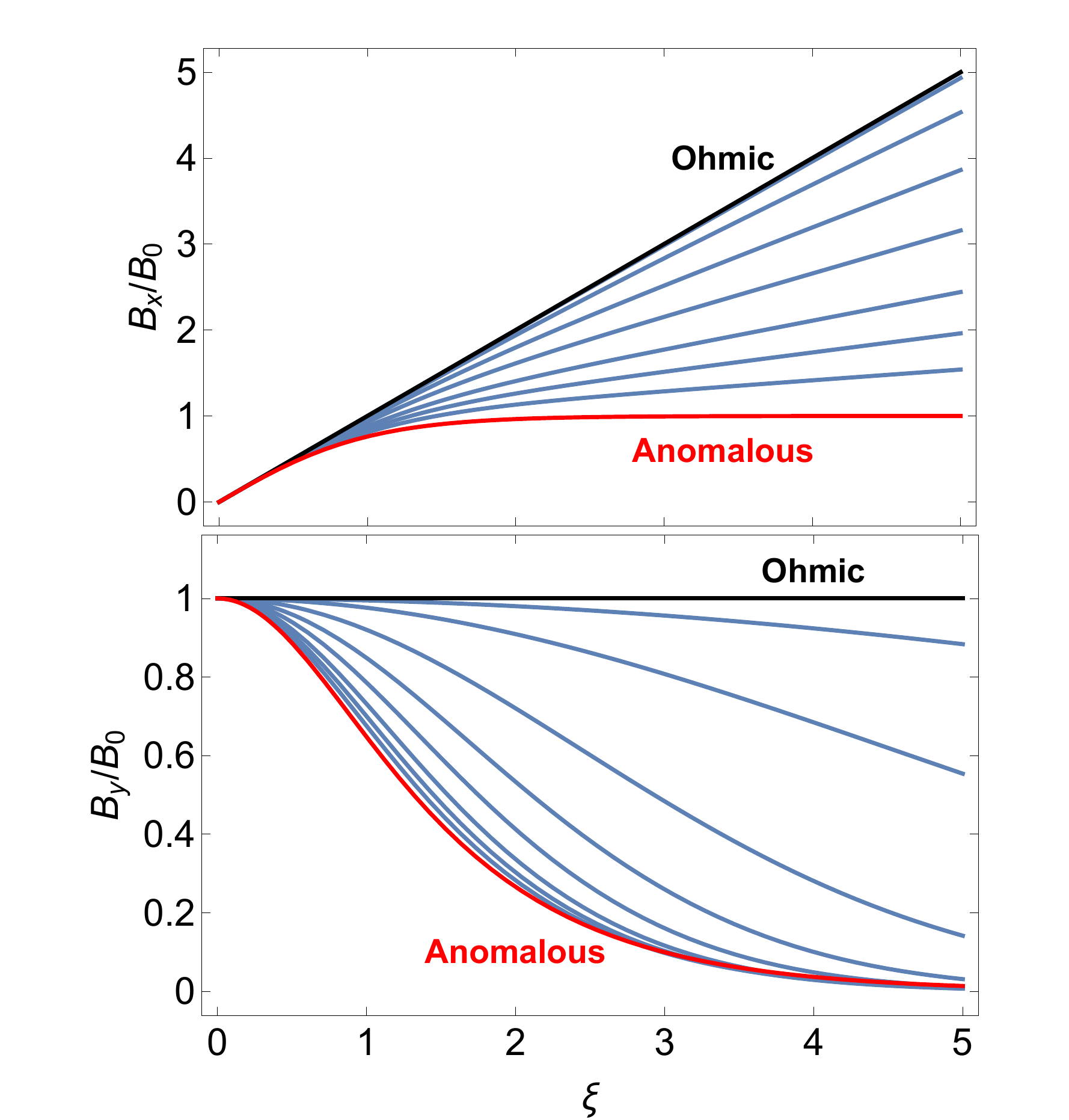}
\caption{\label{fig:BxBy} Components of $\vec{B}$ inside the material for $\beta=0$ (Ohmic current only),0.01,0.05,0.2,0.5,1,2,3.5,7,$\infty$ (anomalous current only) and with $\varphi_0=\pi/2$.}
\end{figure}

We have found that both the Ohmic and anomalous currents generate a component of $\vec{B}$ perpendicular to $\vec{E}$ which grows with distance into the bulk of the material. In the case of the anomalous current however, this growth saturates, and the component of $\vec{B}$ parallel to $\vec{E}$ is simultaneously suppressed over the length scale $\Lambda$. When both currents are nonzero, it is necessary to solve Eq.~\eqref{heqn} numerically; the results are shown in Fig.~\ref{fig:BxBy} for various values of $\beta$. It is evident from the figure that $B_y$ always decays when the anomalous current is present, albeit over longer and longer distances as $\sigma_0$ is increased. 

We also present a solution to Eq.~\eqref{cmeDirac} for the case of a long cylinder (radius $R$) in applied fields $B_0\hat z$ and $E_0\hat z$. Writing $B_z^{in}=B_0\Lambda/[rf(r)]$ where again $\Lambda=(\mu_0\sigma_aE_0B_0)^{-1}$, this equation reduces to
\be
r^2ff''+rff'-r^2(f')^2-1=0,\label{feqn}
\ee
and $B_\phi^{in}=B_0\Lambda(1/r+f'/f)$, $B_r=0$. Eq.~\eqref{feqn} can be solved exactly: $f=\tfrac{1}{2}(k/r+r/k)$, where $k$ is one integration constant, while the other has been chosen to ensure that $B_z$, $B_\phi$ are nonsingular at $r=0$. Since $B_z=B_0$ everywhere outside the cylinder, continuity of $\vec{B}$ requires $f(R)=\Lambda/R$, implying $k=\Lambda+\sqrt{\Lambda^2-R^2}$. The magnetic field inside and outside is then
\bea
B_\phi&=&\frac{2B_0\Lambda r}{r^2+k^2}\Theta(R-r)+\frac{B_0(2\Lambda-k)}{r}\Theta(r-R),\nn\\
B_z&=&\frac{2B_0\Lambda k}{r^2+k^2}\Theta(R-r)+B_0\Theta(r-R).
\eea
This solution is valid for $R<\Lambda$ and assumes that $R$ is sufficiently small that screening of the electric field is negligible. As for the pure CME (Eq.~\eqref{cylsol}), the axial anomaly produces a maximal $B_z$ at $r=0$ and a nonzero $B_\phi$ outside the cylinder with a $\Lambda$-dependent magnitude.

\begin{comment}
Finally, let us write the most general time-dependent equation for this particular model of $\vec{j}$.
We use Maxwell's equations with $\vec{j}=\sigma_a (\vec{E}\cdot\vec{B})\vec{B}+\sigma_0\vec{E}$
with zero charge density. We obtain
$
\sigma_0  \frac{\partial \vec{B}}{\partial t}= \nabla^2 \vec{B} - \frac{\partial^2 \vec{B}}{\partial t^2} 
+ \sigma_a [ \nabla(\vec{E} \cdot \vec{B}) \times \vec{B} + (\vec{E} \cdot \vec{B}) \nabla \times \vec{B} ].
$
Note that for $\sigma_a =0$, we recover the usual equation, which in the quasi-static case describes magnetic diffusion and the skin effect. 
In the time-independent case, we obtain the static equations leading to the 
results discussed in the preceding paragraphs, if $\vec{E}$ is assumed constant. 
\end{comment}

The characteristic length $\Lambda$ can be evaluated using \cite{Son_PRB13,Burkov_PRL14,Zhang_NC16,Li_NC16,LiNatPhys2016,Xiong_Science15}
\be
\sigma_a = l \frac{e^4 \tau_a}{4 \pi^4 \hbar^4 g(E_F)} \label{sigmaa}
\ee
where $l=1,2,...$ is the number of Weyl node pairs, $g(E_F)$ is the density of states at the Fermi level $E_F$ and
$\tau_a$ is the relaxation time for charge pumping between Weyl node pairs. Characteristic values for the Weyl semimetal TaAs \cite{Zhang_NC16} are $\tau_a \approx 5.96 \times 10^{-11} \mathrm{s}$ and
$g(E_F) \approx 10^{41} \mathrm{J^{-1} m^{-3}}$, yielding (for $l = 1$) $\sigma_a \approx 8.25 \times 10^{6} $ $\mathrm{\Omega^{-1} m^{-1} T^{-2}}$. These values yield $\Lambda=(\mu_0\sigma_a E_0B_0)^{-1} \approx$ 9.6 cm, for $E_0$ = 1 V/m and $B_0$ = 1 T as typical experimental values
of applied $\vec{E}$ and $\vec{B}$. A $\Lambda$ of this magnitude would lead to detectable effects in magnetometry measurements performed in SQUID or VSM systems with planar gradiometer geometries to access spatial variations of $\vec{B}$. As an additional diagnostic the magnetometer results can be compared on samples of different sizes, above and below $\Lambda$. An increase in $E_0$ and $B_0$ would result in a proportional decrease in the characteristic length $\Lambda$, allowing a measure of tuning $\Lambda$ to sample size and yet a further diagnostic. Magnetometry hence functions for Weyl fermion materials as an alternative to magnetoresistance measurements, in parallel to the case of superconducting materials. $\Lambda$ can vary widely for the different materials hosting Weyl fermions currently described in the literature, due to differences in $g(E_F)$ and $\tau_a$. In Dirac materials, $g(E_F)=g_s\,g_v\,E_F^2/((2\pi^2)(\hbar v_F)^3)$, where $g_s$ and $g_v$ are the spin- and valley degeneracies \textit{resp.}, and $v_F$ is the velocity constant characterizing the Dirac dispersion. Equation \eqref{sigmaa} then shows that a low $E_F$ (hence low carrier density), high $v_F$ and long $\tau_a$ lead to shorter $\Lambda$ and more readily observable effects in magnetometry. A longer $\tau_a$ is expected to arise from higher carrier mobility \cite{Li_NC16,Xiong_Science15,Xiong_arxiv15b}. Currently TaAs forms a promising candidate, while estimates of $\Lambda$ can be much longer and would not be conducive to ready observations, at least currently, in Cd$_3$As$_2$ microwires \cite{Li_NC16}, or in Zr Te$_5$ \cite{LiNatPhys2016}. However, strides are being made to lower carrier densities and further increase mobilities in many Weyl fermion materials \cite{Xiong_Science15,LiangNatMater2015,Wang_PRB16,Sekhar_NP15}, which will lead to much reduced $\Lambda$.

In conclusion, we have presented a diagnostic procedure for the chiral magnetic effect, additional to negative magnetoresistance and motivated by the fundamental equations of axion electrodynamics. The procedure is based on the penetration of magnetic fields over characteristic length scales in Weyl semimetals.

{\bf Acknowledgements:}
We would like to thank R. Buniy, T. Kephart, G. Sharma and E. N. Economou for interesting conversations and helpful comments.
The work of D.M. was supported in part by the U.S. Department of Energy under contract
DE-FG02-13ER41917. The work of J.J.H. was supported by the U.S. Department of Energy, Office of Basic Energy Sciences, 
Division of Materials Sciences and Engineering under award DOE DE-FG02-08ER46532. 

%\bibliography{refs_quant_EB}

%\end{document}

%

\end{document}